\documentclass[a4paper]{jpconf}
\usepackage{graphicx}
\begin{document}

\title{Azimuthal asymmetries in the unpolarized SIDIS cross section at COMPASS}

\author{G Sbrizzai$^1$}

\address{$^1$ Trieste University and INFN, Physics Department, via Valerio 2 34127, IT}

\ead{giulio.sbrizzai@ts.infn.it}

\begin{abstract}
The study of the spin structure of the nucleon and of the effects rising from the quarks transverse momentum are part of the scientific program of COMPASS, a fixed target experiment at the CERN SPS. The azimuthal asymmetries which appear in the cross-section of SIDIS off unpolarized targets can give insights on the intrinsic momentum structure of the nucleon and on the possible correlation between transverse spin and transverse momentum of the quarks. Here we present the new results for these asymmetries obtained from the COMPASS data collected with a 160 GeV/c positive muon beam impinging on a $^6LiD$ target. The asymmetries are measured for both positive and negative hadrons, and their dependence on several kinematical variable has been studied.
\end{abstract}

\section{Introduction}
\begin{figure}[h]
\includegraphics[width=18pc]{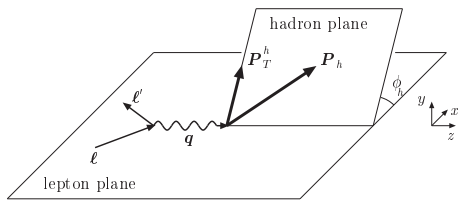}\hspace{2pc}%
\begin{minipage}[b]{10pc}\caption{\label{fig:gns} \small{SIDIS in the Gamma Nucleon System: $\textbf{P}^h$ is the momentum of the produced hadron and $\textbf{P}_T^h$ its transverse component with respect to the virtual photon direction.}}
\end{minipage}
\end{figure}

The SIDIS cross section for longitudinally polarized leptons over unpolarized targets can be expressed~\cite{Bacchetta} as 
\begin{eqnarray}
\frac{\textrm{d}\sigma}{\textrm{d}x\,\textrm{d}y\, \textrm{d}z\, 
\textrm{d}\phi_h\,P_T^h\,\textrm{d}P_T^h}\, &=&
\frac{\alpha^2}{x y Q^2} \frac{(1+(1-y)^2)}{2} \cdot A_0 \Big( 1 + \epsilon _1(y) A^{UU}_{\cos\phi_h} \cos \phi_h  + \nonumber \\
&& + \epsilon _2(y) A^{UU}_{\cos 2\phi_h} \cos2\phi_h + \lambda_l \epsilon _3(y) A^{LU}_{\sin\phi_h} \sin\phi_h   \Big)  \;,
\label{eq:cross_section_1}
\end{eqnarray}
where $A_0 = F_{UU}$ and $A^{UU}_{\cos\phi_h} = \frac{F_{UU}^{\cos\phi_h}}{F_{UU}}$ of reference~\cite{Bacchetta}. 
The angle $\phi_h$ is the angle between the hadron production plane and the lepton plane, as described in figure~\ref{fig:gns}, $\lambda_l$ is the longitudinal polarization of the incident lepton and the quantities $\epsilon _i$ are given by:
\begin{eqnarray}
\epsilon _1 = \frac{2\cdot (2-y)\sqrt{1-y} } {1+(1-y)^2 } , \; \;
\epsilon _2 = \frac{2\cdot (1-y)} {1+(1-y)^2 } , \; \;
\epsilon _3 = \frac {2\cdot y \sqrt{1-y} }{1+(1-y)^2} . 
\end{eqnarray}
The scaling variable is $x = \frac{Q^2}{P\cdot q}$, where $P$ is the target nucleon 4-vector, $q = l - l^{'}$ is the 4 momentum transfer (with $l$ and $l^{'}$ the 4 momenta of the incoming and scattered lepton) and $Q^2 = -q^2$. $y=\frac{P\cdot q}{P \cdot l}$ is the fraction of energy of the incoming nucleon transferred to the target nucleon. 

The $\sin\phi_h$ modulation has no clear interpretation in terms of Parton Model. Its amplitude is proportional to the beam polarization $\lambda _l$ and it has been found to be small but different from zero ($5\%$ with a proton target) by the CLAS collaboration ~\cite{clas1} which used a $6$ GeV electron beam. 

The amplitudes of the $\cos\phi_h$ and the $\cos2\phi_h$ modulations have been measured in unpolarized SIDIS on a proton target by previous experiments and results have been published by  EMC ~\cite{EMC_1} and E665 ~\cite{E665} in a kinematical region similar to COMPASS and at higher energies by the ZEUS experiment ~\cite{ZEUS}. More recent results, for positive and negative hadrons separately, have been presented by HERMES~\cite{hermes} and CLAS~\cite{mher}. The origin of these two modulations is related with the so called
Cahn effect, with the Boer-Mulders Transverse Momentum Dependent Parton Distribution Function (TMD PDF) and with the perturbative QCD corrections, which are expected to be important only for hadron transverse momentum $P_T^h > 1$ GeV/c. 

The contribution to the $\cos\phi_h$ and the $\cos2\phi_h$ azimuthal modulation given by the Cahn effect comes from kinematics. It is calculated starting from the elastic quark-lepton cross section ~\cite{Cahn}, taking into account the intrinsic transverse momentum of the quark inside the nucleon $\textbf{k}_T$. The resulting $\cos\phi_h$ amplitude is proportional to $<k_T>/Q$ and the $\cos2\phi_h$ amplitude is proportional to $<k_T^2>/Q^2$. Extraction of the mean value of $k^2_T$ from the measured amplitude of the $\cos\phi_h$ modulation has been already presented ~\cite{Anselmino_1}.

The Boer-Mulders TMD PDF, which describes the correlation between transverse momentum and transverse polarization of a quark inside an unpolarized nucleon, has been introduced recently ~\cite{BM}. It contributes to the $\cos\phi_h$ and the $\cos2\phi_h$ azimuthal modulation in unpolarized SIDIS where it appears convoluted with the Collins Fragmentation Function (FF). In some model calculations ~\cite{Barone} it is predicted that the contribution of the Boer-Mulders function to the $\cos2\phi_h$ asymmetry is comparable with the Cahn effect and that it significantly differs for positive or negative hadrons. A first extraction of the Boer-Mulders PDF from previous data has already been preformed~\cite{Melis}.

New results on these azimuthal asymmetries have been produced by the COMPASS experiment and they are shown here for the first time. 

\section{Experiment}
The COMPASS experiment ~\cite{COMPASS} is a fixed target experiment at SPS at CERN. It uses a $160$ GeV/$c$ $\mu^+$ beam, with a natural polarization of $-80\%$, and a solid state polarized target. From 2002 to 2006, data has been taken with a polarized $^6$LiD target.
The targets were either transversely or longitudinally polarized to accomplish the wide experimental program on the spin structure of the nucleons. 
Up to 2004 the target consisted of two separate cells filled with the same material, oppositely polarized. The polarization of the cells have been reversed every week, during the data taking periods, in order to minimize systematic effects.

\section{Analysis}
The data used to extract the results are the part of the 2004 run in which the target was transversely polarized. To extract the amplitudes of the azimuthal modulations present in equation~\ref{eq:cross_section_1}, the data with opposite polarization have been combined to cancel the small spin dependent contributions. Only the events with the $\mu^+$ scattered in the downstream cell have been used in order to have a larger angular acceptance for the detected hadrons. 

Apart for the almost standard DIS cuts: $Q^2 > 1$ GeV$^2$, $0.003 < x < 0.13$, $0.1 < y < 0.9$, $W > 5 $GeV / $c^2$ (where $W$ is the invariant mass of the $\gamma ^* P$ system), a cut on the transverse hadron momentum has been introduced $0.1 < P_T ^h < 1.0$ GeV/$c$ where the lower limit is chosen in order to have a good resolution on the azimuthal angle, while the upper limit ensures negligible pQCD corrections. A cut on $z=E_h / E_{\gamma^*}$ with $0.2 < z < 0.85$ is applied to reject hadrons from the target fragmentation (lower limit) and away from possible diffractive effects at $z=1$. Finally, cuts on the virtual photon angle (calculated with respect to the nominal beam direction), asking $\theta_{\gamma ^*} ^{lab} < 60$ mrad, and $y>0.2$ are applied for a better hadron acceptance. 

Dedicated Monte Carlo simulations have been performed to correct the hadron azimuthal distributions for the apparatus acceptance. For each of these simulations: the SIDIS event generation is performed with LEPTO; the interaction between particles and the experimental apparatus together with the detectors response is simulated with COMGEANT, a software based on GEANT 3; the Monte Carlo data are finally reconstructed with the same software (CORAL) used for the real data. 

\begin{figure}[h]
\begin{minipage}{25pc}
\includegraphics[width=25pc]{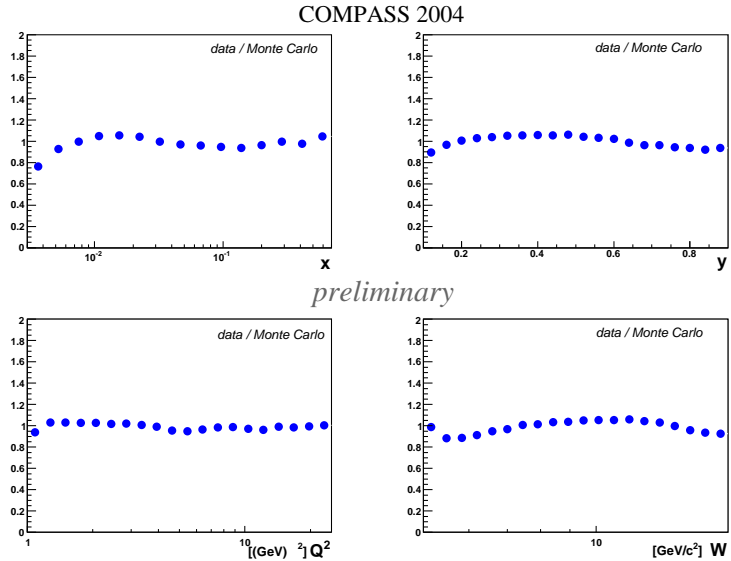}
\end{minipage}
\begin{minipage}{10pc}
\caption{\label{fig:mc_dis} \small{Ratios between kinematical distributions obtained from real data and Monte Carlo data for the DIS variables: $x$, $y$, $Q^2$ and $W$.}}
\end{minipage}
\end{figure}

\begin{figure}
\begin{minipage}{25pc}
\includegraphics[width=25pc]{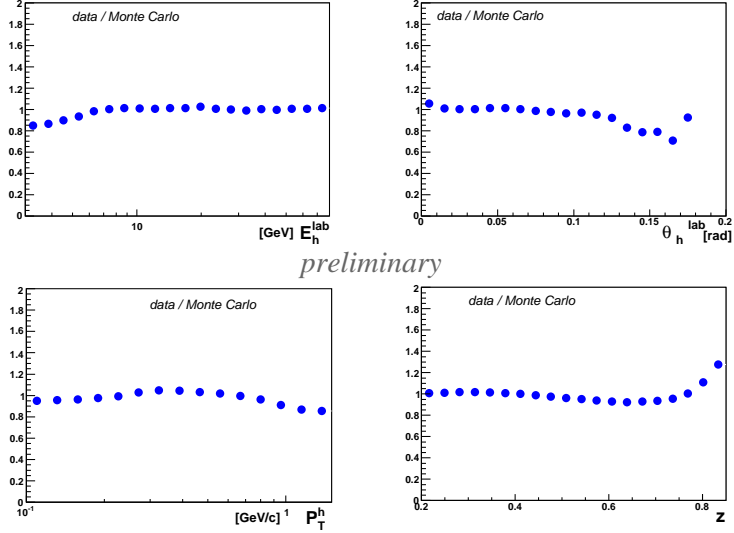}
\end{minipage}
\begin{minipage}{10pc}
\caption{\label{fig:mc_had} \small{Ratios between kinematical distributions obtained from real data and Monte Carlo data for the hadron variables: the energy $E^h_{lab}$ and the polar angle $\theta_{lab}^h$, both calculated in the Laboratory System, $P_T^h$ and $z$.}}
\end{minipage} 
\end{figure}

Before evaluating the effects of the apparatus acceptance, extensive studies have been performed comparing the kinematical distributions of the reconstructed real and Monte Carlo data, in order to improve the agreement but limiting as much as possible the tuning of the LEPTO parameters. The ratios data/Monte Carlo for the adopted setting of the LEPTO parameters and after the standard DIS cuts are shown in figure~\ref{fig:mc_dis} for the DIS variables and in figure~\ref{fig:mc_had} for the hadron variables. The agreement is good enough to correct the azimuthal distributions and to get the final results.

The azimuthal acceptance has been evaluated as: $Acc(\phi_h) \,=\, N_{rec}(\phi^{rec}_h) / N_{gen}(\phi^{gen}_h)$, where $N_{rec}(\phi_h)$ is the reconstructed hadron distribution (after all analysis cuts) and $N_{gen}(\phi_h)$ is the generated hadron distribution, obtained from Monte Carlo simulations.

The hadron azimuthal distribution corrected for the apparatus acceptance $N_{corr}(\phi_h)\,=\, N(\phi_h) / Acc(\phi_h)$ has been fitted, binning alternatively in $x$, $z$ and $P_T^h$, for positive and negative hadrons separately, with a four parameter function: $N_{corr}(\phi_h)\,=\, N_0 \cdot (1\,+\,p_{\cos(\phi_h)}\cdot \cos(\phi_h)\,+\,p_{\cos(2 \phi_h)}\cdot \cos(2 \phi_h)\,+\,p_{\sin(\phi_h)}\cdot \sin(\phi_h))$ where the free parameters $p_i$ are the amplitudes of the three independent azimuthal modulation ($p_{\cos\phi_h} = \epsilon_1 A^{UU}_{\cos\phi_h}$). 

\begin{figure}[h]
\begin{minipage}{15pc}
\includegraphics[width=13pc]{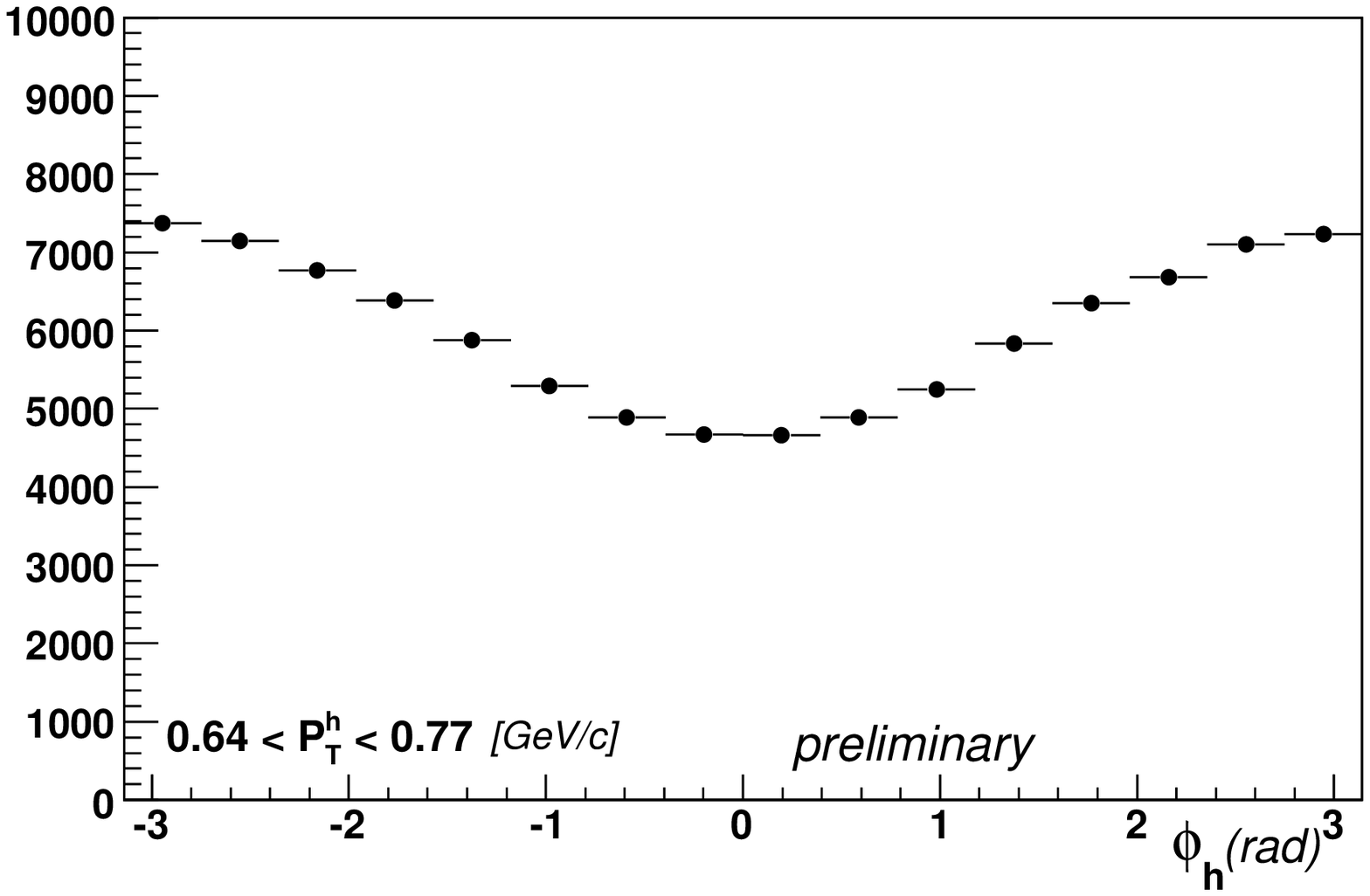}
\caption{\label{fig:hnnn} Measured azimuthal \\ distribution $N(\phi_h)$.}
\end{minipage} 
\begin{minipage}{13pc}
\includegraphics[width=13pc]{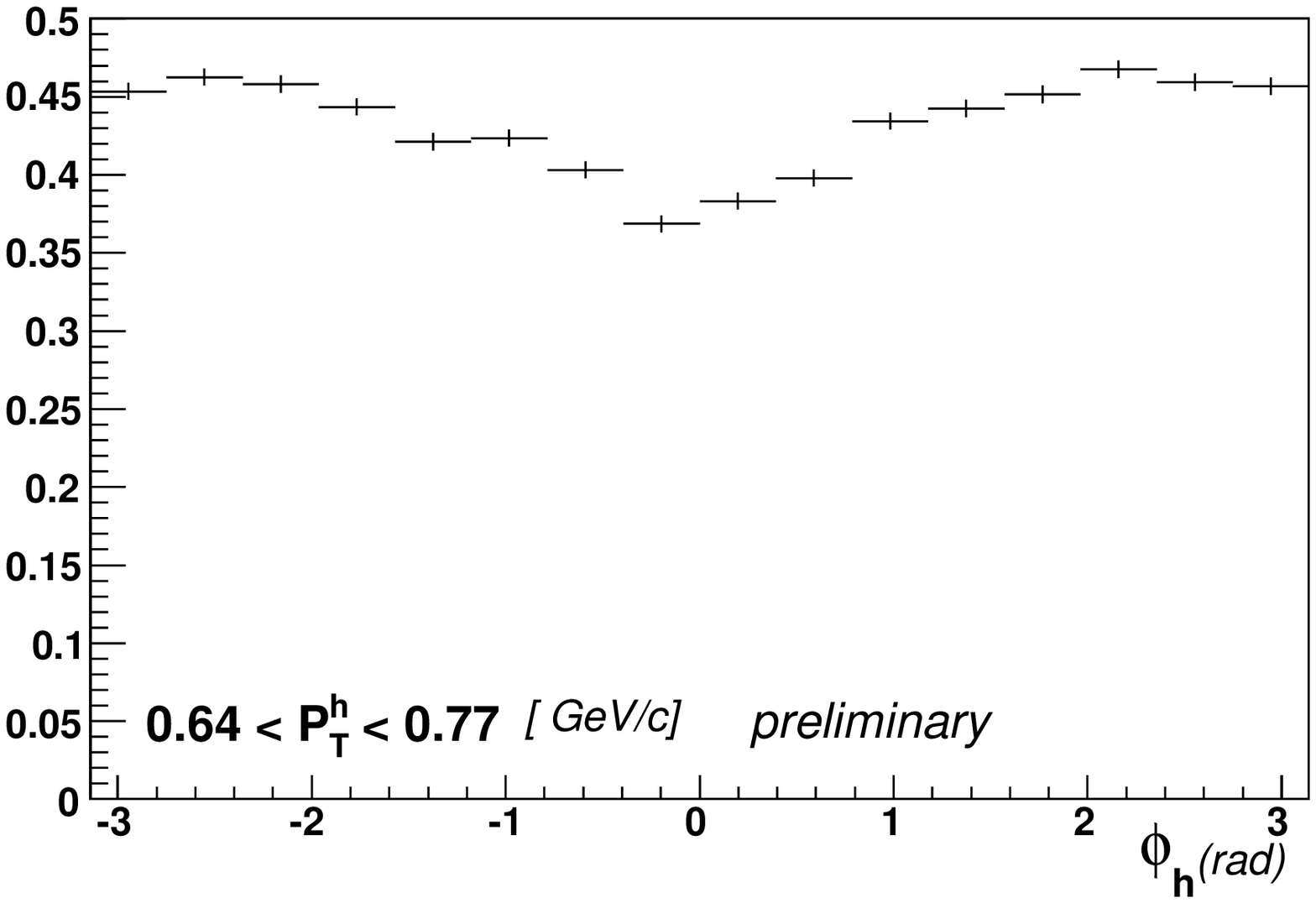}
\caption{\label{fig:hacc} Azimuthal acceptance $Acc(\phi_h)$.}
\end{minipage}\\
\begin{minipage}{15pc}
\includegraphics[width=13pc]{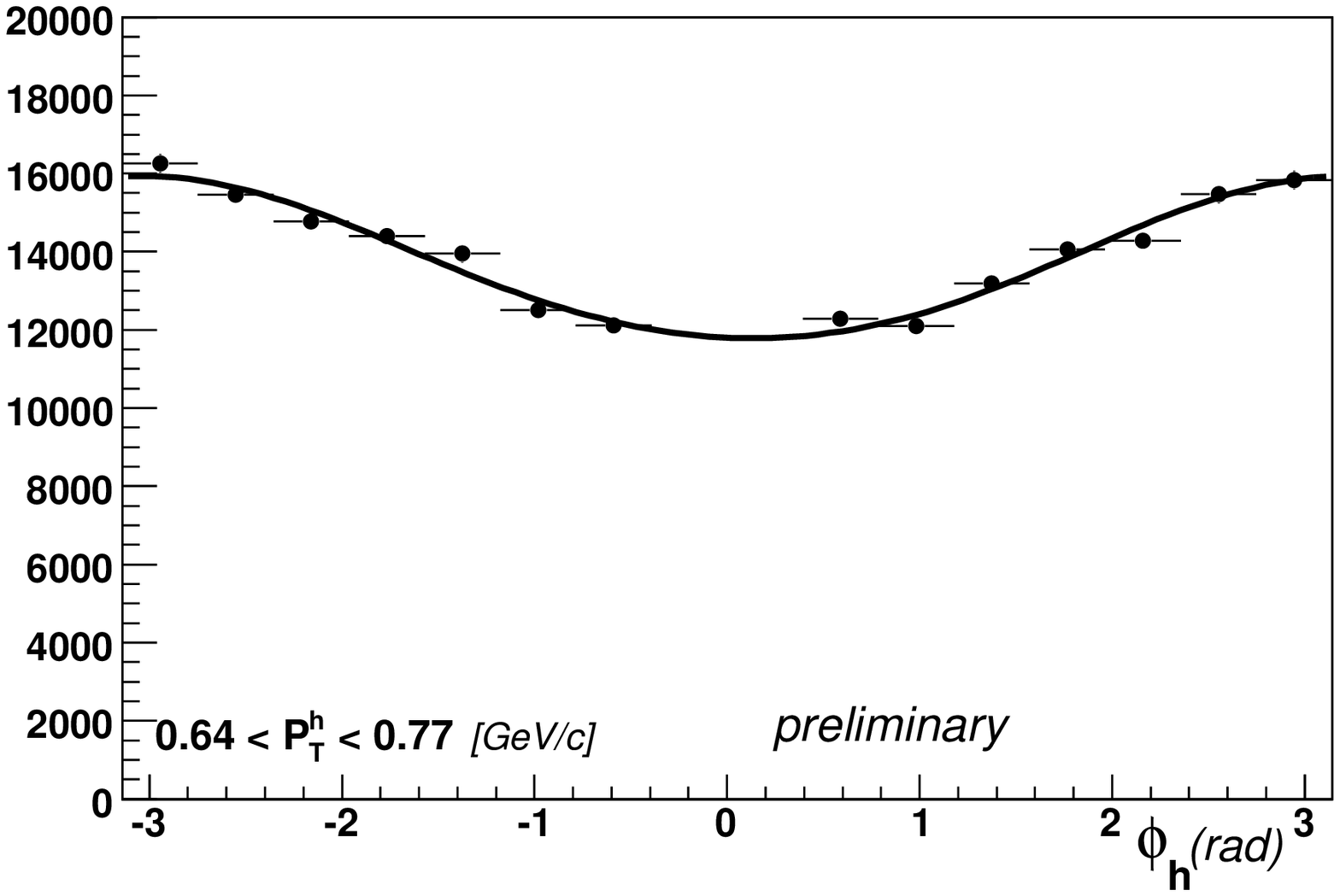}
\end{minipage}
\begin{minipage}{16pc}
\caption{\label{fig:hcor} Measured azimuthal distribution corrected by the acceptance $N_{corr}$.}
\end{minipage}
\end{figure}

A typical hadron azimuthal distribution from real data $N(\phi_h)$, the corresponding acceptance from the Monte Carlo simulation $Acc(\phi_h)$ and the corrected distribution $N_{corr}(\phi_h)$ are shown in figure~\ref{fig:hnnn}, figure~\ref{fig:hacc} and figure~\ref{fig:hcor} respectively.

\begin{figure}[h]
\includegraphics[width=20pc]{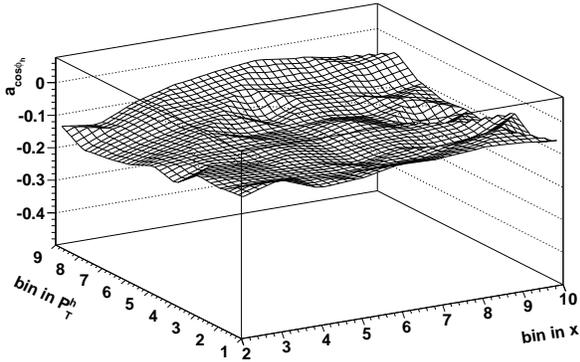}\hspace{2pc}
\begin{minipage}[b]{12pc}\caption{\label{fig:acc} \small{The amplitude $a_{\cos\phi_h}(x, P_T ^h)$ of the $\cos\phi_h$ modulation extracted from the fit to $A(\phi_h) \,=\, N_{rec}(\phi_h) / N_{gen}(\phi_h)$ calculated in each 2 dimensional bin in $x$ and $P_T ^h$. The absolute value of the amplitude is less than $15\%$ over the whole kinematical range and it is does not show any partcular trend.}}
\end{minipage}
\end{figure}

The amplitudes of the azimuthal modulations introduced by the acceptance have been studied at length in the given kinematics. 
The azimuthal acceptances have been calculated in 2 dimensions where one variable is chosen among $x$, $z$ or $P_T ^h$ and the other among a set of other valuables on which the acceptance could have a relevant dependence like $x$, $z$, $P_T ^h$, $y$, $\theta_{\gamma ^*} ^{lab}$ and $Q^2$. The kinematical regions in which the corrections are large have been identified and excluded; the most relevant cuts $\theta_{\gamma ^*} ^{lab} < 0.06 rad$ and $y > 0.2$. The overall acceptance turned out to be flat. As an example the amplitude $a_{\cos\phi_h}(x, P_T ^h)$ of the $\cos\phi_h$ modulation extracted from a fit to $Acc(\phi_h) \,=\, N_{rec}(\phi_h) / N_{gen}(\phi_h)$ calculated in each 2 dimensional bin in $x$ and $P_T ^h$ is shown in figure~\ref{fig:acc}. 
Similar pictures have been obtained studying other azimuthal modulations and for the combinations of the other kinematical variables. Thus the method used to extract the asymmetries should give the same results both using different event generators and with acceptances extracted in multi dimensional bins.

\section{Systematic studies}
\begin{figure}
\begin{minipage}{27pc}
\includegraphics[width=27pc]{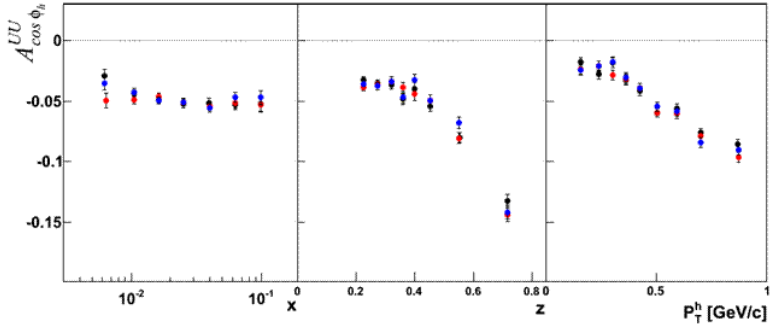}
\end{minipage}
\begin{minipage}{10pc}
\caption{\label{fig:sys3mc} \small{Comparison between the $A^{UU}_{\cos\phi_h}$ azimuthal asymmetries extracted using 3 Monte Carlo samples using different event generations. The blue points are calculated with the Monte Carlo simulation used to extract the final results.}}
\end{minipage}
\end{figure}

The acceptance has been also calculated using two other Monte Carlo sets of data produced by changing the PDFs set and the LEPTO parameters relative to the transverse momentum distribution in the fragmentation process and the Lund Fragmentation Function. The agreement between the real data and these two sets of Monte Carlo data is worse than that of figure~\ref{fig:mc_dis} and figure~\ref{fig:mc_had}, and shows opposite trends in the $x$, $y$ and $\theta_{lab}^h$ distributions. For this reason they have been used to check the whole procedure. As an example of this test, in figure~\ref{fig:sys3mc}, $A^{UU}_{\cos\phi_h}$ from positive hadrons extracted using the 3 different Monte Carlo is shown. As can be seen the differences are small as expected if the azimuthal acceptance is really flat in the kinematical variables.

The difference of the asymmetries extracted using the 3 Monte Carlo has been used to evaluate the systematic error. To the systematic error contributes also the difference between the results obtained from the data taken with the transversely polarized target and the one obtained from the data taken with the longitudinally polarized target. In this second case the configuration of the apparatus is different, some detectors have been moved and many parameters changed and so, for the longitudinal target polarization data, a different Monte Carlo sample has been used to extract the unpolarized asymmetries. The effects due to possible inefficiencies of some detector planes have been also evaluated and found to be smaller than the previous ones. The sum in quadrature of all these three different contributions has been used to calculate the systematic error, which is twice the statistical one.

Other tests have been performed finding negligible systematic effects. In particular the radiative correction has been calculated for in the given kinematics using RADGEN and are relatively small. The amplitudes of non physical modulations have been added in the fit and turned out to be compatible with zero.

\section{Results}

\begin{figure}
\begin{minipage}{30pc}
\includegraphics[width=30pc]{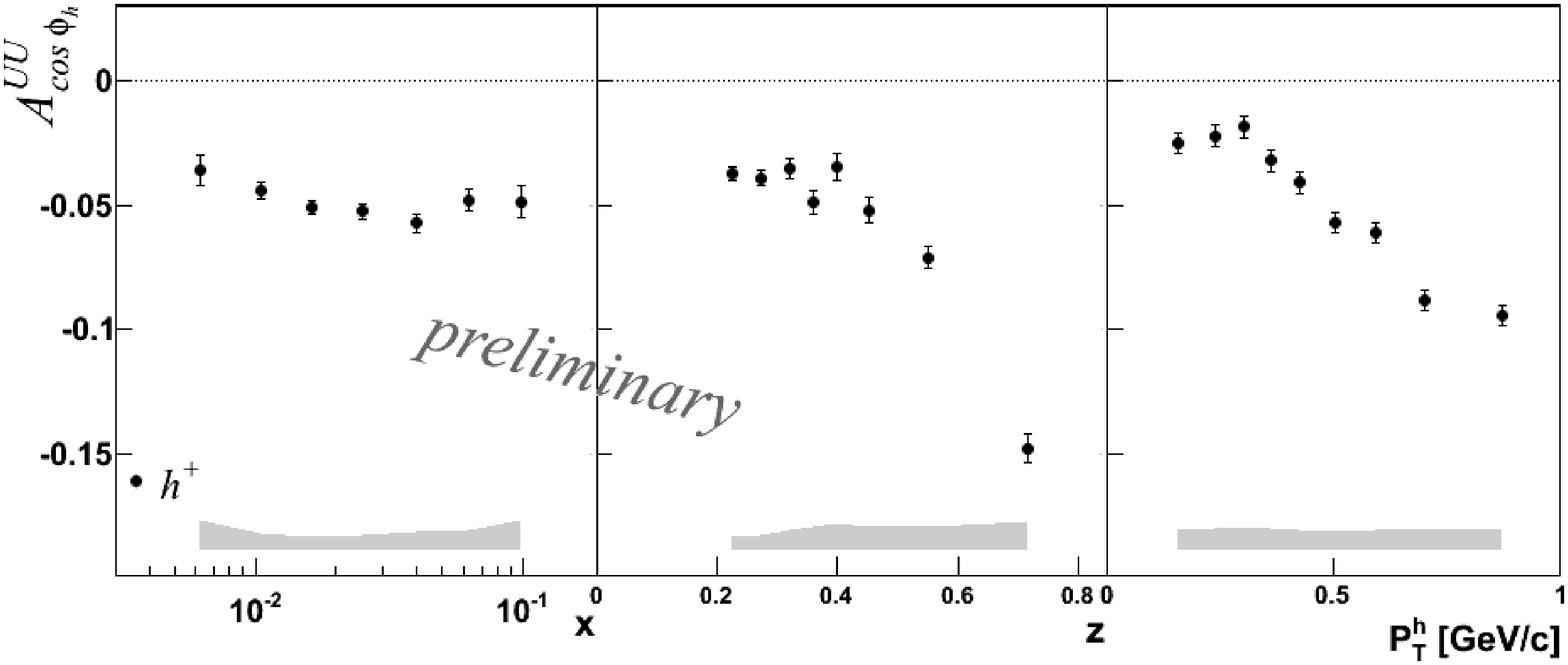}\\
\includegraphics[width=30pc]{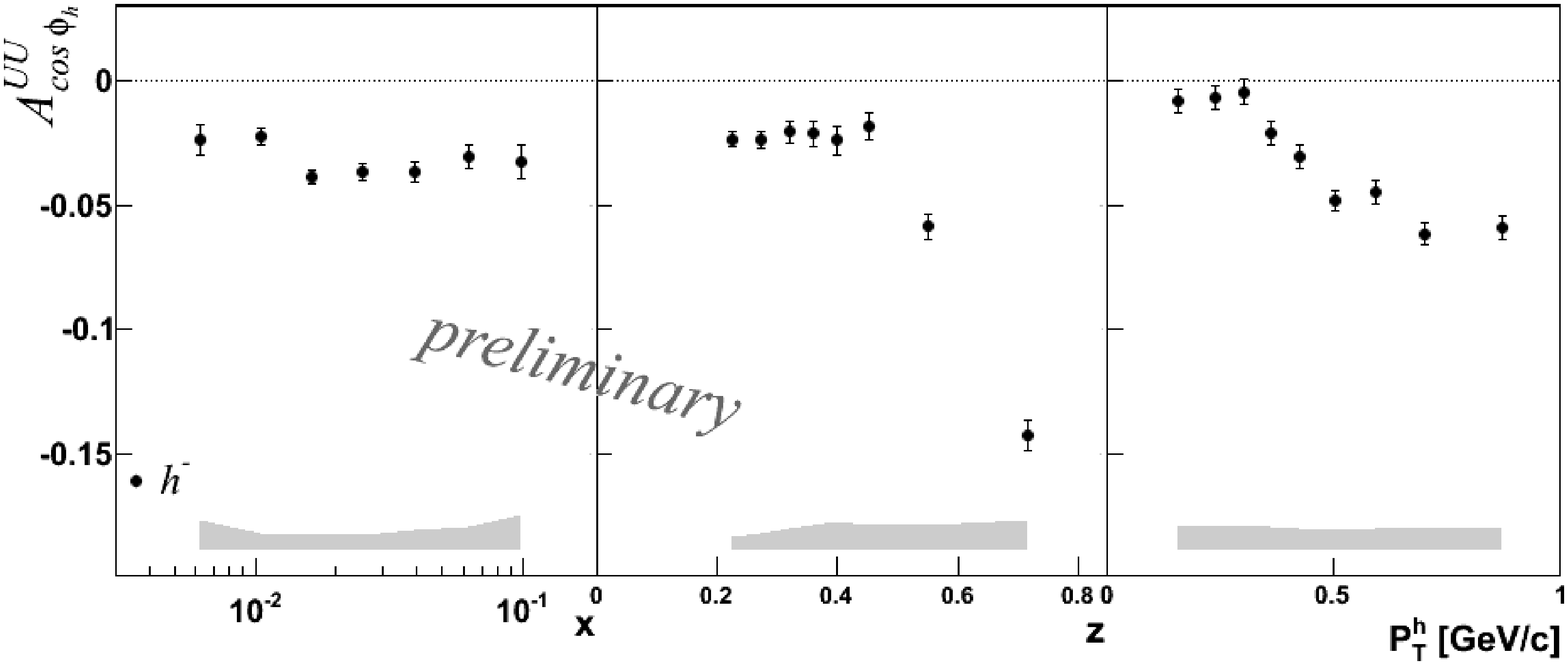}
\end{minipage}
\begin{minipage}{7pc}
\caption{\label{fig:cos} \small{Results for the $A^{UU}_{\cos\phi_h}$ azimuthal asymmetries as functions of $x$, $z$ and $P_T ^h$, obtained for positive hadrons (upper plot) and negative hadrons (lower plot). Grey bands are the systematic error \\($\sigma _{sys} = 2 \cdot \sigma _{stat}$).  }}
\end{minipage}
\end{figure}

\begin{figure}
\begin{minipage}{30pc}
\includegraphics[width=30pc]{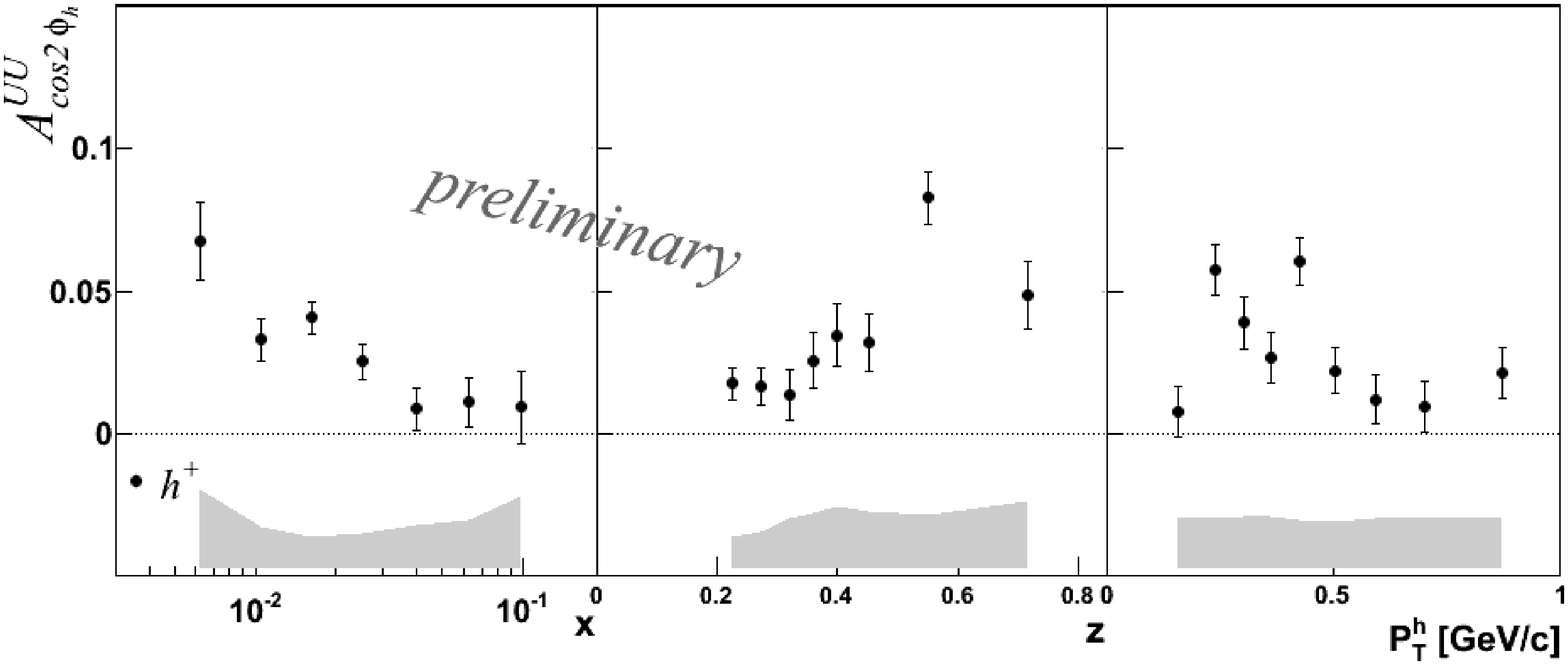}\\
\includegraphics[width=30pc]{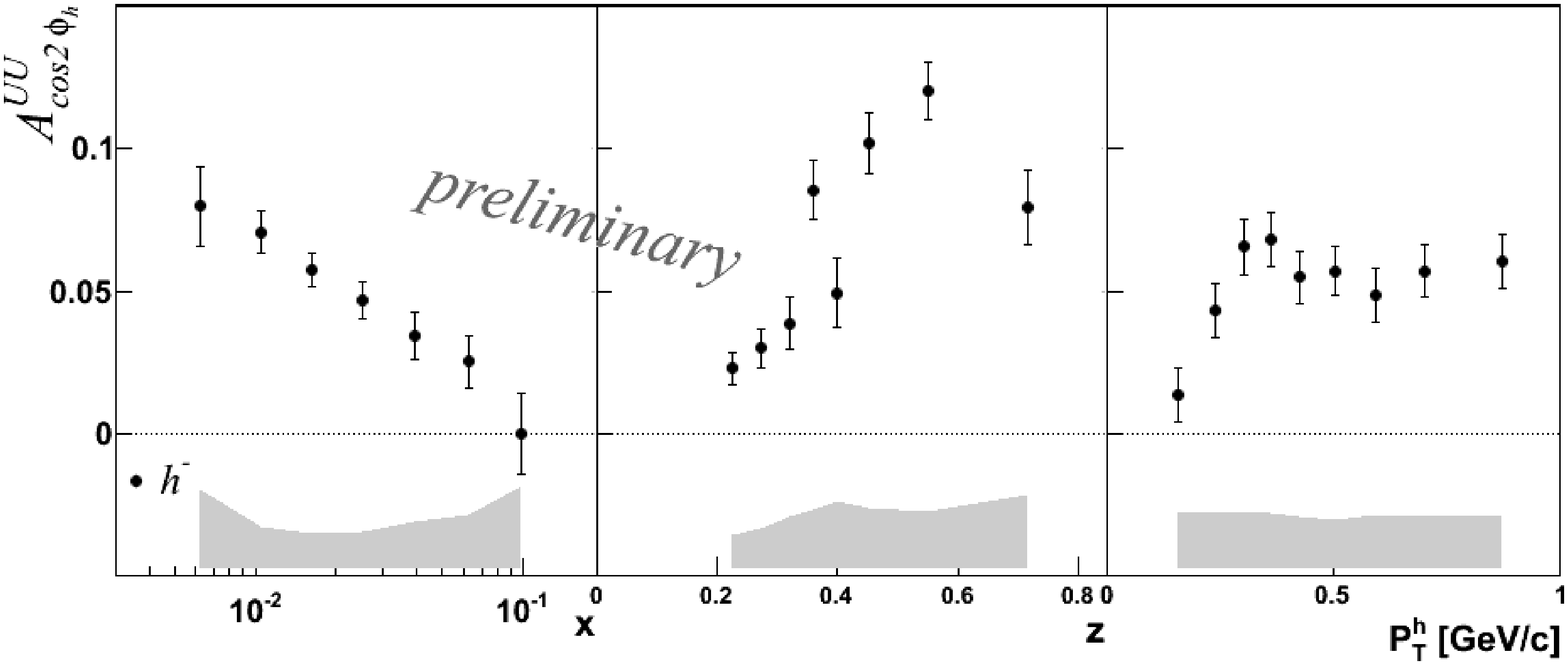}
\end{minipage}
\begin{minipage}{7pc}
\caption{\label{fig:cos2} \small{Results for the $A^{UU}_{\cos2\phi_h}$ azimuthal asymmetries as functions of $x$, $z$ and $P_T ^h$, obtained for positive hadrons (upper plot) and negative hadrons (lower plot). Grey bands are the systematic error \\($\sigma _{sys} = 2 \cdot \sigma _{stat}$). }}
\end{minipage}
\end{figure}

\begin{figure}
\begin{minipage}{30pc}
\includegraphics[width=30pc]{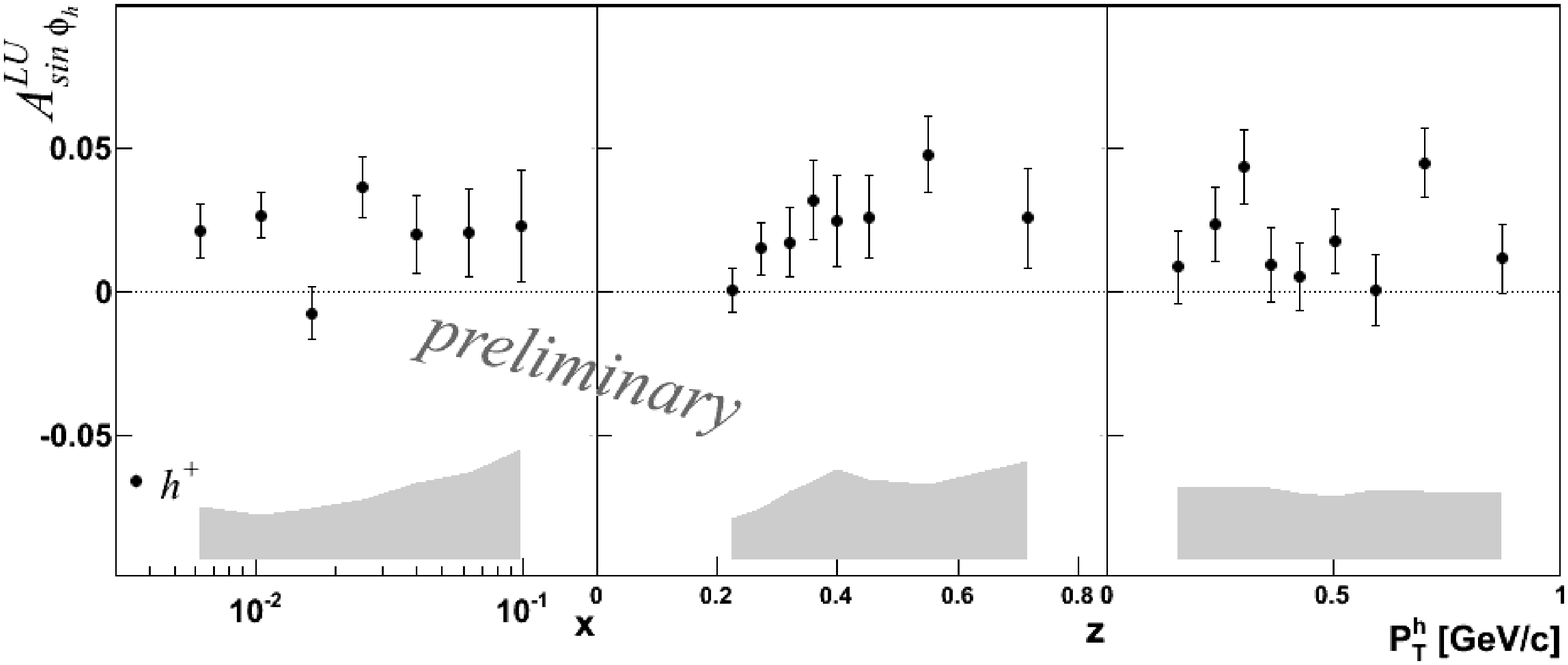}\\
\includegraphics[width=30pc]{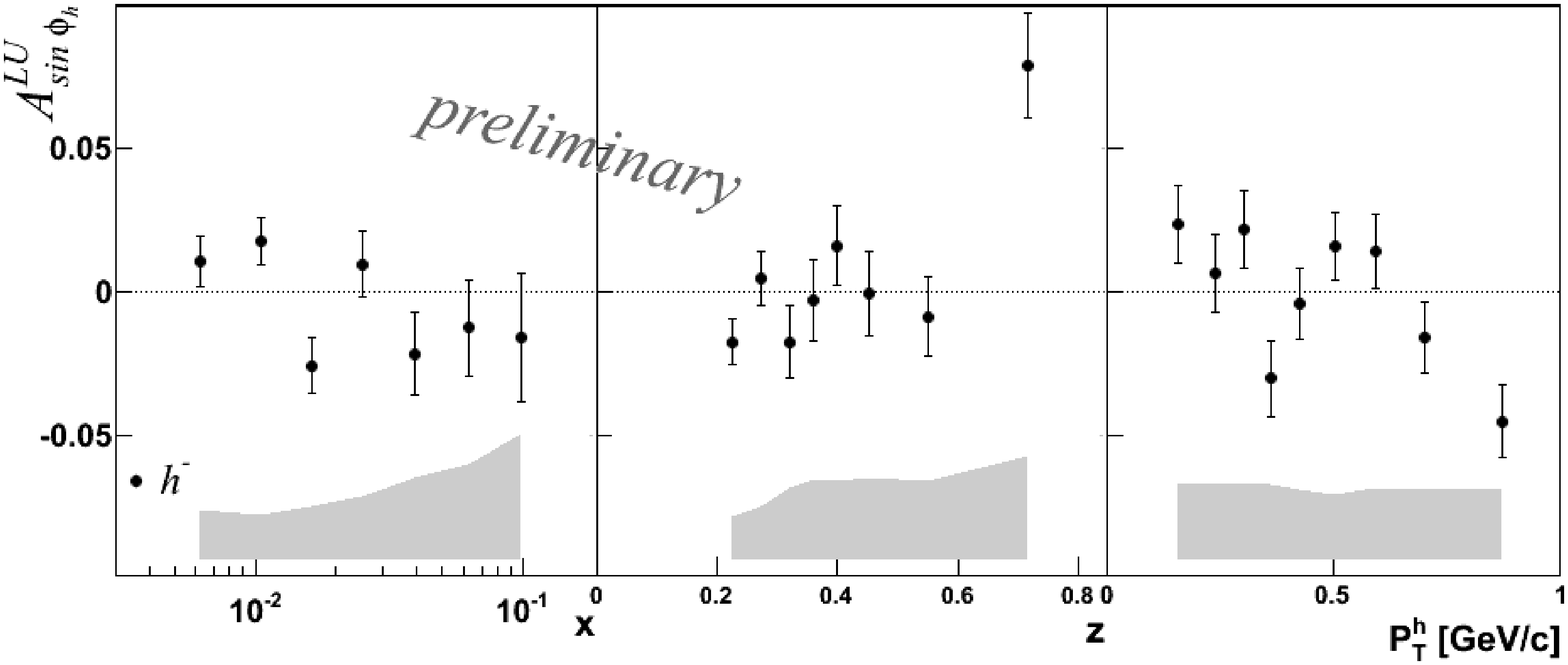}
\end{minipage}
\begin{minipage}{7pc}
\caption{\label{fig:sin} \small{Results for the $A^{LU}_{\sin\phi_h}$ azimuthal asymmetries as functions of $x$, $z$ and $P_T ^h$, obtained for positive hadrons (upper plot) and negative hadrons (lower plot). Grey bands are the systematic error \\($\sigma _{sys} = 2 \cdot \sigma _{stat}$). }}
\end{minipage}
\end{figure}

The results for the azimuthal asymmetries extracted from the COMPASS $^6LiD$ data are shown in figure~\ref{fig:cos} for the $A^{UU}_{\cos\phi_h}$ amplitude, in figure~\ref{fig:cos2} for the $A^{UU}_{\cos2\phi_h}$ amplitude, \clearpage
and in figure~\ref{fig:sin} for the $A^{LU}_{\sin\phi_h}$ amplitude. They are shown separately for positive (top) and negative (bottom) hadrons and are given as functions of $x$, $z$ or $P_T ^h$.

\begin{figure}
\begin{minipage}{27pc}
\includegraphics[width=27pc]{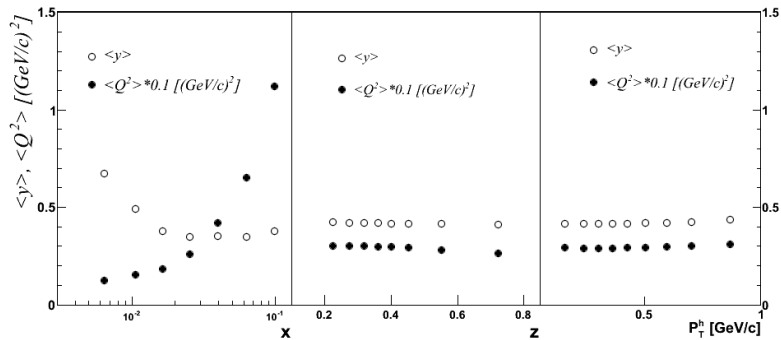}
\end{minipage}
\begin{minipage}{10pc}
\caption{\label{fig:means} \small{$<y>$ and $<Q^2>$ mean values as functions of $x$, $z$ and $P_T ^h$. The $<Q^2>$ mean value has been multiplied by a factor $10^{-1}$.}}
\end{minipage}
\end{figure}

The largest signal is given by the $A^{UU}_{\cos\phi_h}$ amplitude: up to $15\%$ and with a strong dependence both on $z$ and $P_T^h$. The dependence on $x$ is less evident and its interpretation should take into account the correlation between $x$ and $Q^2$, shown in figure~\ref{fig:means}. The $A^{UU}_{\cos\phi_h}$ asymmetry should be mainly due to the Cahn effect and the unexpected dependence of the results from the charge of the produced hadron could hint both to a different value of the intrinsic transverse momentum $k_T$ for the $u$ and the $d$ quarks or to the  Boer-Mulders TMD PDF. This function appears here as an higher twist effect while comes at leading twist in the $A^{UU}_{\cos2\phi_h}$ amplitude, thus is expected to give a relevant contribution to this last asymmetry. 

The $A^{UU}_{\cos2\phi_h}$ asymmetry is smaller showing a strong trend as function of all the three studied kinematical variables $x$, $z$ or $P_T ^h$. The dependence of $A^{UU}_{\cos2\phi_h}$ on the hadron charge is clearly seen and could be related to the Boer-Mulders function and thus to an important role of the correlation between the intrinsic transverse momentum and spin inside the nucleon. 

A clear hadron charge dependence has been measured also by the HERMES Collaboration~\cite{hermes} both for the $A^{UU}_{\cos\phi_h}$ and the $A^{UU}_{\cos2\phi_h}$ amplitudes. Any comparison with the COMPASS results should take into account also for the different kinematical range covered by the two experiments. 

The $A^{LU}_{\sin\phi_h}$ azimuthal asymmetries show a positive signal for the positive hadrons while are compatible with zero for the negative ones. A signal of the same sign has been measured by the CLAS Collaboration ~\cite{mher} using a $6$ GeV electron beam on a proton target. 

Summarizing, the COMPASS SIDIS measurements with unpolarized targets show large asymmetries and effects sometimes unexpected. Comparison with model calculations and the outcome of the other experiments will allow to clarify the present description of the nucleon structure including the role of the intrinsic transverse momentum and its correlation with the transverse spin.

\clearpage

\clearpage

\section*{References}
\begin{thebibliography}{11}

\bibitem{Bacchetta}
A.Bacchetta, M.Diel, K.Goebe, A.Metz, P.Mulders and M.Schlegel, 
JHEP \textbf{0702}, (2007) 093


\bibitem{clas1}
CLAS Collaboration, H.~Avakian et~al. 
Phys. Rev. {\bf D69} (2004) 112004.

\bibitem{EMC_1}
European Muon Collaboration, J.J.Aubert et al. 
Phys. Lett. B \textbf{130}, (1983) 118.


\bibitem{E665}
E665 Collaboration, M.R.Adams et al.
Phys. Rev. D \textbf{48} (1993) 5057-5066.


\bibitem{ZEUS}
ZEUS Collaboration, J.Breitweg et al.
Phys. Lett. B \textbf{481} (2000) 199-212.


\bibitem{hermes}
F. Giordano, on behalf of the HERMES Collaboration,
\emph{These proceedings}.





\bibitem{Cahn}
R.N.Cahn, 
Phys. Lett. B \textbf{78} 269, (1978).

\bibitem{Anselmino_1}
M.Anselmino, M. Boglione, U, D'Alesio, A.Kotzinian, F.Murgia and A.Prokudin,
Phys. Rev. D \textbf{71}, (2005) 074006.


\bibitem{BM}
D.Boer, P.J.Mulders
Phys. Rev. D \textbf{57} (1998) 5780-5786.

\bibitem{Barone}
V.Barone, A.Prokudin and B.Q.Ma
Phys. Rev. D \textbf{78} (2008) 045022. 

\bibitem{Melis}
V.Barone, S.Melis and A.Prokudin
Phys. Rev. D\textbf{81} (2010) 114026

\bibitem{COMPASS}
COMPASS Collaboration, P.Abbon et al.
Nucl. Instr. and Meth. A \textbf{577} (2007) 455-518


\bibitem{mher}
M. Aghasyan, on behalf of the CLAS Collaboration,
\emph{These proceedings}.

\end{thebibliography}
\end{document}